\documentclass{article}

\usepackage{amsmath}
\usepackage{graphicx}
\usepackage{caption}
\usepackage{subcaption}

\renewcommand{\Re}{\operatorname{Re}}
\renewcommand{\Im}{\operatorname{Im}}

\begin{document}

\title{Load impedance of immersed layers
       on the quartz crystal microbalance:
       a comparison with colloidal suspensions of spheres}

\author{M. Mel\'endez\footnote{marc.melendez@uam.es}, A. V\'azquez-Quesada, R. Delgado-Buscalioni}

\maketitle

\begin{abstract}
The analytical theories derived here for the acoustic load impedance measured
by a  quartz crystal microbalance (QCM), due to  the presence of  layers
of different  types (rigid, elastic and viscous) immersed in  a fluid,
display generic properties, such as ``vanishing  mass'' and positive
frequency  shifts, which have been  observed  in QCM  experiments  with
soft-matter systems.  These phenomena seem  to contradict the well-known
Sauerbrey relation at the heart of many QCM measurements, but here
we show that they arise as a natural consequence of hydrodynamics.
We compare our one-dimensional immersed plate theory with three-dimensional
simulations of rigid and flexible sub-micron-sized suspended spheres, and
with experimental results for adsorbed micron-sized colloids which yield
a ``negative acoustic mass''. The parallel behaviour unveiled indicates that
the QCM response is highly sensitive to hydrodynamics, even for adsorbed colloids. Our conclusions call for a revision of existing theories
based on adhesion forces and elastic stiffness at contact, which should
in most cases include hydrodynamics.
\end{abstract}


\section{Introduction}

Quartz crystal microbalances (QCM) principally consist of a thin quartz crystal
between two electrodes. Being a piezoelectric material, the quartz oscillates in
response to an AC current. In many devices, the crystal is cut in such a way
that transverse vibrations take place parallel to the free surface. Due to the
oscillatory motion, we would reasonably expect (correctly, as it turns out) that
increasing the mass slightly by adding a small layer onto the surface of the QCM
will lead to a small decrease in the frequency of oscillation, $f$. For a
harmonic oscillator of mass $M$, we can easily prove to ourselves that a small
increase in the mass $\Delta M$ causes a decrease in frequency
\begin{equation}
  \frac{\Delta f}{f}= -\frac{1}{2}\frac{\Delta M}{M}
\end{equation}
In 1959, Sauerbrey derived a similar equation for the QCM \cite{Sauerbrey1959}.
With $\Delta m$ representing mass per unit area deposited on the QCM,
\begin{equation}
 \frac{\Delta f}{f} = -\frac{2f_0 \Delta m}{Z_Q}.
\end{equation}
Typically, for AT-cut crystals the fundamental frequency equals
$f_0 = 5\ \mathrm{MHz}$, the acoustic impedance of the quartz
$Z_Q = 8.8 \times 10^6\ \mathrm{kg}/(\mathrm{m}^2\mathrm{s})$, and $f$
represents the working frequency. As frequency shifts may be measured with great
accuracy, much experimental work has relied on the QCM as an extremely
sensitive mass detector, hence its name.

QCMs also work in contact with fluids, as demonstrated by Nomura and Okuhara
\cite{Nomura1982}, who showed that the transverse waves propagate into a fluid
(of density $\rho$ and shear viscosity $\eta$) with a heavy damping. Thus, QCM
with dissipation monitoring (QCM-D) instruments measure both the frequency of
oscillation and the energy dissipation in a ring-down experiment, in which the
AC voltage is turned off and the quartz crystal allowed to come to rest. Because
the damping occurs within tens of microseconds, a series of consecutive
ring-downs can monitor the evolution of molecular processes taking place over
second or minute time scales. As a consequence, QCM-D instruments have become a
standard tool for biosensing systems of supported membranes, and
Langmuir-Blodgett, protein and liposome films, among others
\cite{Johannsmann2015,Johannsmann2008,Lane2011,Voinova2002}. In this context,
Gizeli's group observed that, for some systems, the ratio of the dissipation to
the frequency shift, which they termed the \textit{acoustic ratio}, did not
depend on the concentration of molecules deposited on the QCM, suggesting that
it was a property of the geometry of the molecules, rather than the mass of the
deposited film \cite{Johannsmann2015}.

When thinking in terms of the Sauerbrey relation, one would never conceive of an
increase in the load leading to an increased frequency of vibration, but this has
in fact been observed in experiments with massive (micron-sized) particles
\cite{Dybwad1985,Marxer2003,Pomorska2010,Kravchenko2019}. A widely repeated explanation
for these ``negative Sauerbrey masses'' states that an increase in the frequency shift
(.i.e a ``negative acoustic mass'') arises as a consequence of a very fast response
of the analyte-wall contact, modelled by a (generally complex) effective spring
\cite{Johannsmann2015}. However, this explanation does not take into account the
hydrodynamic transport of momentum. Historically, hydrodynamic effects have often been
disregarded in QCM research. As we shall see below, however, the acoustic ratio may
well diverge or change sign naturally \textit{even for moderately small loads
suspended in a fluid}.

Although previous research has developed one-dimensional phenomenological models of
the viscoelasticity of films \cite{Voinova2002}, recent experiments with
nanoparticles, liposomes, viruses and DNA strands have shown strong deviations
from these theories
\cite{Johannsmann2015,Johannsmann2008b,Johannsmann2009,Reviakine2011,Tsortos2008}. 

Following most of the previous work in the field, we make use of the small load
approximation \cite{Johannsmann2008}, which allows us to write the complex
frequency shift in terms of the load impedance.
\begin{equation}
  \label{small_load_approximation}
  \Delta f  + \mathrm{i} \Delta \Gamma = \frac{\mathrm{i} f_0}{\pi} \frac{Z_L}{Z_Q}.
\end{equation}
Here, $\Delta \Gamma$ is the change in the decay rate of the resonator and the
complex load impedance $Z_L$ equals the stress phasor on the QCM surface divided by
its velocity phasor.

The main point in the present article is that the primary effect determining the
impedance of suspensions measured by the QCM involves the change in the
hydrodynamic motion of the fluid due to the presence of suspended matter. To
argue for this statement, we derived an analytical theory for the effect of an
infinite immersed layer on the motion of a QCM, represented by a flat horizontal
oscillating plane at $z = 0$ in contact with a fluid filling the space $z > 0$.
We also show that the changes in impedance as a function of distance and
frequency for a suspended membrane resemble the changes observed for a sparse
periodic array of spheres. The data for spheres was produced by our QCM
simulations of suspended liposomes using the FLUAM code \cite{Balboa2012}, which
is based on Peskin's immersed boundary method \cite{Peskin2002}. We have shown
elsewhere that our simulations agree with experimental results
\cite{Tsortos2020,Delgado2020}. In addition, we have considered the crossover to
positive frequency shift (``negative acoustic mass''). Comparing the analytical
prediction of the plate system with our simulations of sub-micron spheres and
experiments carried out with micron-sized colloids leads to several interesting
conclusions, discussed below.

\section{Oscillating boundary layer}
We wish to study fluid sytems near a vibrating plane wall in the Stokes flow
regime \cite{Stokes1851}. Our fluid, with density $\rho$ and shear viscosity
$\eta$, obeys the equations of linear hydrodynamics and lies in the space above
the $z = 0$ plane. Furthermore, the translational symmetry of the problem
ensures that the flow velocity $\tilde{u}(z, t)$ will depend only on the $z$ coordinate
and time. The equation describing the velocity field reads
\cite{Stokes1851,Kanazawa1985,Landau1987}
\begin{equation}
  \frac{\partial \tilde{u}}{\partial t} = \frac{\eta}{\rho} \frac{\partial^2 \tilde{u}}{\partial z^2}.
\end{equation}
We will mark time-dependent oscillating functions with tildes and rely on phasors
to represent the amplitudes of the steady-state solutions,
\begin{equation}
  \tilde{u}(z, t) = \Re\left(u(z) e^{-\mathrm{i} \omega t}\right),
\end{equation}
with the complex-valued phasor amplitude $u(z)$ \cite{Kanazawa1985},
\begin{equation}
  \label{Stokes_boundary_layer}
  u(z) = A e^{-\alpha z} + B e^{\alpha z},
\end{equation}
where $\alpha = (1 - \mathrm{i})/\delta$, and $\delta = \sqrt{2 \eta/(\omega \rho)}$
measures the depth of penetration of the oscillating flow.
Assume the wall vibrates with frequency $\omega$. If we apply stick boundary
conditions where the fluid meets the plane and add that the velocity vanishes
far from it, then
\begin{align}
  u(0) & = u_0, \\
  \lim_{z \to \infty} u(z) & = 0.
\end{align}
The coefficient $u_0$ may take, in general, complex values. By applying the
boundary conditions to the general solution, we see that $A = u_0$ and $B = 0$,
which implies a velocity phasor
\begin{equation}
  \label{unperturbed_Stokes_flow}
  u_f(z) = u_0 \exp(-\alpha z)
\end{equation}
for the Stokes flow.

We now calculate the impedance associated to the Stokes flow, $Z_f$,
According to the standard definition, $Z_f$ is the ratio of the shear stress
exerted on the plane to its velocity. Let $\sigma = \eta \left.\frac{\partial u}{\partial z}\right|_{z = 0}$ represent the phasor amplitude of the stress.
\begin{equation}
  Z_f = \frac{\sigma}{u(0)}
      = \frac{\eta \left.\frac{\partial u}{\partial z}\right|_{z = 0}}{u_0}
      = -\eta \alpha.
\end{equation}
The subscript $f$ distinguishes the impedance of the base Stokes flow from
other load impedances calculated below.

\section{\label{immersed_rigid_plate}Immersed rigid plate}

\begin{figure}
  \begin{center}
    \begin{subfigure}[b]{.485\linewidth}
      \includegraphics[width = \linewidth]{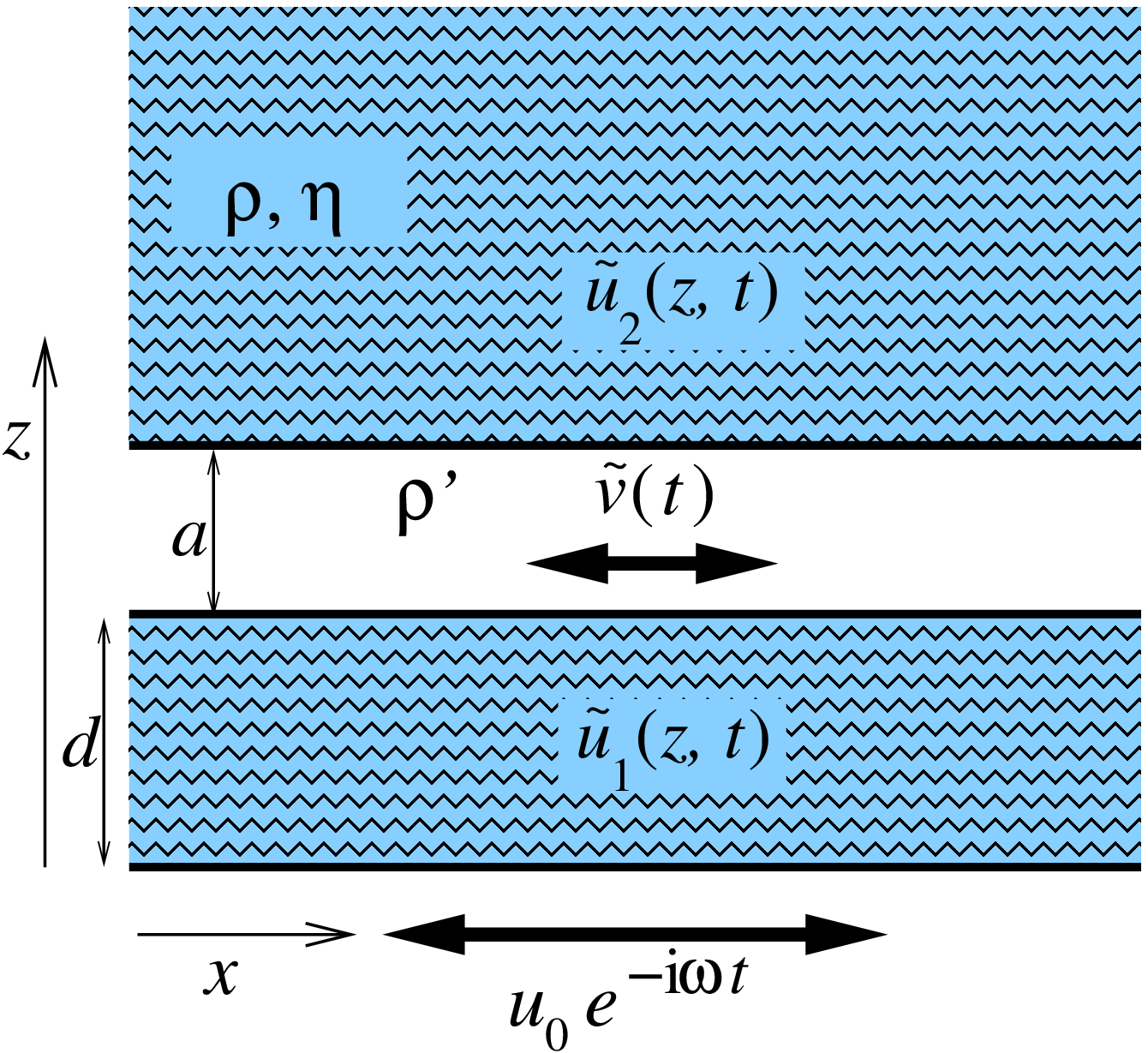}
      \caption{\label{system_diagram}}
    \end{subfigure}
    \begin{subfigure}[b]{.485\linewidth}
      \includegraphics[width = \linewidth]{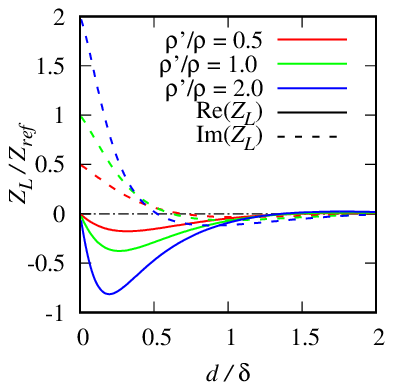}
      \caption{\label{impedance_vs_height}}
    \end{subfigure}
    \begin{subfigure}[b]{.465\linewidth}
      \includegraphics[width = \linewidth]{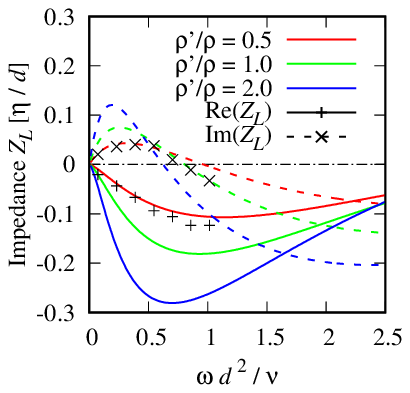}
      \caption{\label{impedance_vs_freq}}
    \end{subfigure}    
    \begin{subfigure}[b]{.525\linewidth}
      \includegraphics[width = \linewidth]{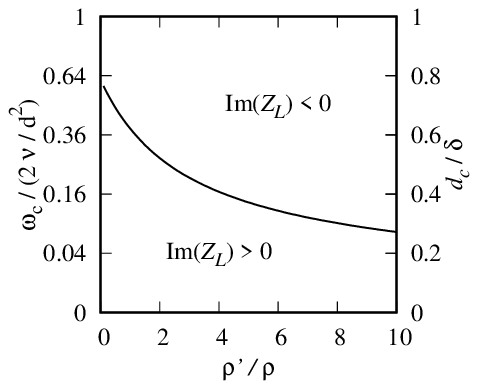}
      \caption{\label{zcut_vs_rhop}}
    \end{subfigure}
  \end{center}
  \caption{(\subref{system_diagram}) Schematic representation of a rigid
           horizontal plate of density $\rho'$ and thickness $a$
           immersed in a fluid of density $\rho$ and shear viscosity $\eta$
           above an oscillating plane.
           (\subref{impedance_vs_height}) Complex impedance $Z_L$ due to the
           presence of the plate in Fig. \ref{system_diagram} versus distance
           $d$ between plate and plane in units of
           $\delta = \sqrt{2\nu/\omega}$  (the kinematic viscosity equals
           $\nu = \eta/\rho$). The curves correspond to the real (solid) and
           imaginary (dashed) parts of the impedance (divided by the Sauerbrey
           value $Z_{ref}=\omega \rho^{\prime} a$).
           (\subref{impedance_vs_freq}) Load impedance versus dimensionless
           parameter $\omega d^2 / \nu$ for three different plate densities.
           The solid plots $\Re(Z_L)$, while the dashed line corresponds to
           $\Im(Z_L)$. The thickness was chosen equal such that $d/a = 1$. The
           experimental points for silica particles with a radius of half a micron
           in a $150\ \mathrm{mM}$ KCl electrolite were
           taken from Ref. \cite{Olsson2012} (setting $d = 50\ \textrm{nm}$ for
           a qualitative comparison).
           (\subref{zcut_vs_rhop}) Frequency $\omega_c$ and height $d_c$ at
           which the imaginary part of the impedance crosses the horizontal
           axis (see Fig. \ref{impedance_vs_height}) versus the ratio of the
           plate density to the fluid density. Please note that the scale on the
           left is not linear. Eq. (\ref{solid_plate_load_impedance}) implies that
           changing $\rho'/\rho$ is equivalent to changing $a$ while leaving
           $\rho'/\rho$ fixed.
           }
\end{figure}

Now imagine a solid horizontal plate of thickness $a$ and mass density $\rho'$
placed above the vibrating plane at a distance $d$ (Fig. \ref{system_diagram}).
The fluid will transmit the motion of the vibrating lower plane and drag the
suspended layer along. Having reached a stationary oscillation, the plate will
move with velocity
\begin{equation}
  \tilde{v}(t) = v_0 e^{-\mathrm{i} \omega t},
\end{equation}
with a complex factor $v_0$ to be determined below. The motion of the solid layer
results from the shear stress exerted by the fluid from above and below. Let
$\tilde{u}_i(z,t)$ represent the velocity fields in the regions below ($i=1$) and
above ($i=2$) the plate. Then we can rewrite Newton's equation of
motion,
\begin{equation}
  \rho' a \frac{d\tilde{v}}{dt}
    = \eta \left( \left.\frac{\partial \tilde{u}_2}{\partial z}\right|_{z = d + a}
                  - \left.\frac{\partial \tilde{u}_1}{\partial z}\right|_{z = d} \right),
\end{equation}
in terms of phasor amplitudes,
\begin{equation}
  \label{wall_equation_of_motion}
-\mathrm{i} \omega \rho' a v_0
    = \eta \left( \left.\frac{\partial u_2}{\partial z}\right|_{z = d + a}
                  - \left.\frac{\partial u_1}{\partial z}\right|_{z = d} \right).
\end{equation}

Clearly, the fluid above the plate obeys the equations of Stokes flow already
calculated above, but this time with an amplitude given by the motion of the
plate.
\begin{equation}
  u_2(z) = v_0 e^{-\alpha (z - (d + a))}.
\end{equation}
For the lower wall, we substitute the form of the general solution
(\ref{Stokes_boundary_layer}) into boundary conditions which ensure that the
fluid moves at the same speed as the walls at the point of contact.
\begin{align}
  \label{Boundary_conditions}
  u_1(0) & = u_0 = A + B \nonumber \\
  u_1(d) & = v_0 = A e^{-\alpha d} + B e^{\alpha d} 
\end{align}
The extra load impedance due to the plate, $Z_L$, equals the total impedance minus the
impedance due to the base Stokes flow $Z_f$,
\begin{equation}
  Z_L = \frac{\sigma}{u_0} - Z_f
      = 2 \alpha \eta \frac{B}{u_0}.
\end{equation}
From the boundary conditions (\ref{Boundary_conditions}) we obtain $A$ and
$B$ as a function of the plate and resonator velocity amplitudes, $v_0$ and 
$u_0$, and write the load impedance as
\begin{equation}
  \label{load_impedance}
  Z_L = \frac{\alpha \eta}{\sinh(\alpha d)}\ \frac{1}{u_0} \left (v_0-u_f(d)\right).
\end{equation}
Thus, a solid plate creates an impedance proportional to the difference
between the plate velocity $v_0$ and the (unperturbed) Stokes flow velocity
(\ref{unperturbed_Stokes_flow}) at the lower fluid-plate interface ($z = d$). Eq.
(\ref{load_impedance}) leads to the conclusion that a fixed plate ($v_0 = 0$)
yields an impedance inversely proportional to $1-e^{2 \alpha d}$, and that the load
impedance vanishes if the plate moves with the base Stokes flow ($v_0 = u_f(d)$). In
the limiting case of a small gap between vibrating wall and plate, $\alpha d \ll 1$,
the load impedance corresponds to that of a Couette flow created by the perturbative
velocity $v_0 - u_f(d)$ in a gap of width $d$,
\begin{equation}
  Z_L = \eta \left(\frac{v_0-u_f(d)}{d}\right), \ \ \text{for } d \ll \delta.
\end{equation}

All that remains now is to determine $v_0$ as a function of $u_0$. To this end,
we substitute the general solution for Stokes flow (\ref{Stokes_boundary_layer})
into the equation of motion for the wall (\ref{wall_equation_of_motion}) and use
the result in combination with the boundary conditions to solve for $v_0$. Substituting
the result into Eq. (\ref{load_impedance}),
\begin{equation}
  \label{solid_plate_load_impedance}
  Z_L = \frac{\omega \rho' a}
             {\frac{\omega \rho' a}
                   {2 \alpha \eta}
              \left(e^{-2 \alpha d} - 1\right) - \mathrm{i}}\ 
        e^{-2 \alpha d}.
\end{equation}
Note that when the distance between the plate and the lower plane vanishes, we recover
a Sauerbrey-like relation, $\lim_{d \to 0} Z_L = \mathrm{i} \omega \rho' a$, with
a purely imaginary impedance, corresponding to a frequency shift proportional to
the deposited mass $\rho' a$. The opposite limit obviously leads to a vanishing load
impedance, $\lim_{d \to \infty} Z_L = 0$. Below, we will often use the Sauerbrey
impedance, $Z_{ref}=\omega \rho^{\prime} a$, as a reference to scale our results.

\subsection{Diverging and negative acoustic ratios}

Fig. \ref{impedance_vs_height} plots the load impedance as a function of the height
$d$ for three different plate densities. As already mentioned, within the small load
approximation  (\ref{small_load_approximation}), the dimensionless acoustic ratio
(defined as the ratio of the dissipation to the frequency shift) is proportional to
$-2\Re(Z_L)/\Im(Z_L)$. In experimental work, it is customary to present a ratio of
the dissipation $\Delta D$ to the frequency shift $\Delta f$, related to the
dimensionlness acoustic ratio by
\begin{equation}
    -\frac{\Delta D}{\Delta f} = -\frac{2}{f_n}\ \frac{\Re(Z_L)}{\Im(Z_L)},
\end{equation}
where $f_n$ is the frequency of the harmonic used in experiments. Therefore, if the
imaginary part of $Z_L$ changes sign as a consequence of the variation of some parameter,
the acoustic ratio will diverge and become negative after the divergence.

Positive frequency shifts (negative Sauerbrey masses) show up in experiments when
analysing massive particles above a certain crossover frequency $\omega_c$. The simple
plate model also displays such an inversion of the sign of $Z_L$. In particular, Fig.
\ref{impedance_vs_freq} shows that the plate qualitatively behaves like experiments
with micron-sized colloids. The plate load impedance is compared there to measurements of
silica particles of radius $R = 0.5\ \mu\mathrm{m}$ adsorbed to a silica surface in a
$\mathrm{K}^+\mathrm{Cl}^-$ electrolite at a concentration of $150\ \mathrm{mM}$ 
\cite{Olsson2012}. The similarity between particle and plate suggests that the load
impedance results principally from hydrodynamic stress, in contrast to previous research,
which had attributed the effect to contact forces between the surface and the load
\cite{Dybwad1985,Pomorska2010,Olsson2012,Johannsmann2015}. We will return to this
important point below.

Rescaling the load impedance by $Z_{ref}$ leaves us with an expression that depends
only on the dimensionless parameters $\rho'/\rho$, $a/\delta$ and $d/\delta$.
\begin{equation}
  \label{crossover_distance}
  \frac{Z_L}{Z_{ref}} = \frac{e^{-2 (1 - \mathrm{i}) d / \delta}}
                             {\frac{1 + \mathrm{i}}{2} \frac{\rho'}{\rho}
                              \frac{a}{\delta}
                              \left(e^{-2 (1 - \mathrm{i}) d / \delta} - 1\right)
                              - \mathrm{i}}.
\end{equation}
Because $\delta^2 \propto \omega^{-1}$, doubling the layer width $a$ and the distance
$d$ has the same effect as multiplying the frequency $\omega$ by four. Thus, for any
fixed frequency we expect to observe a diverging acoustic ratio ($\Im(Z_L) = 0$) for
some large enough distance $d_c$ (Fig. \ref{zcut_vs_rhop}). Similarly, large enough
analytes ($a > a_c$) yield negative frequency shifts for given values of $\omega$ and
$d$. Setting $\Im(Z_L) = 0$ leads to the following relation among the dimensionless
parameters:
\begin{equation}
  \frac{\rho'}{\rho} \frac{a_c}{\delta_c}
    = \frac{2 \cos\left(2 \frac{d_c}{\delta_c}\right)}
           {e^{-2 d_c/\delta_c}
           \left(
           \cos\left(2 \frac{d_c}{\delta_c}\right)
           - \sin\left(2 \frac{d_c}{\delta_c}\right)
           \right)},
\end{equation}
where we have used the subindex $c$ as a reminder that we mean crossover values. We
will illustrate the generality of the hydrodynamic effect below by comparing this
prediction to simulations of immersed spheres and experiments with colloidal particles.

\subsection{The hydrodynamic origin of ``negative acoustic masses''}
 
\begin{figure}
  \begin{center}
    \begin{subfigure}[b]{.485\linewidth}
      \includegraphics[width = \linewidth]{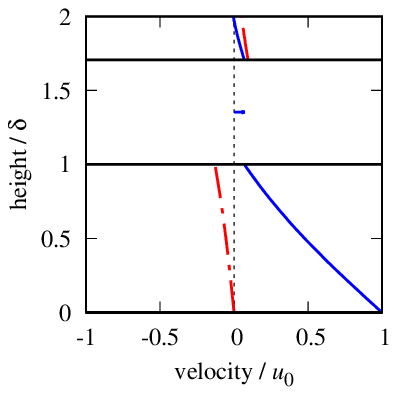}
      \caption{\label{velocity_profile.theta_0.d_1}
               $d = \delta$, $\theta = 0$, $\Re(\sigma) < 0$, $\Delta \Gamma > 0$.}
    \end{subfigure}
    \begin{subfigure}[b]{.485\linewidth}
      \includegraphics[width = \linewidth]{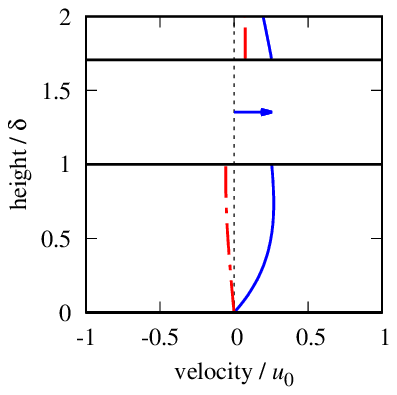}
      \caption{\label{velocity_profile.theta_pio2.d_1}
               $d = \delta$, $\theta = \pi/2$, $\Im(\sigma) < 0$, $\Delta f > 0$.}
    \end{subfigure}
    \begin{subfigure}[b]{.485\linewidth}
      \includegraphics[width = \linewidth]{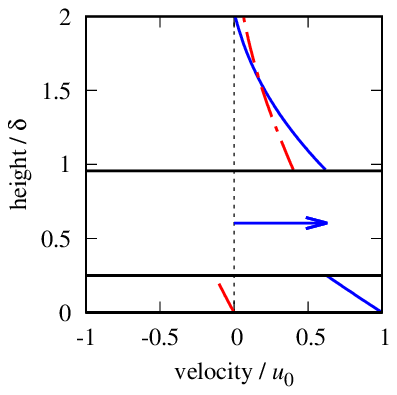}
      \caption{\label{velocity_profile.theta_0.d_0.25}
               $d = \delta/4$, $\theta = 0$, $\Re(\sigma) > 0$, $\Delta \Gamma > 0$}
    \end{subfigure}
    \begin{subfigure}[b]{.485\linewidth}
      \includegraphics[width = \linewidth]{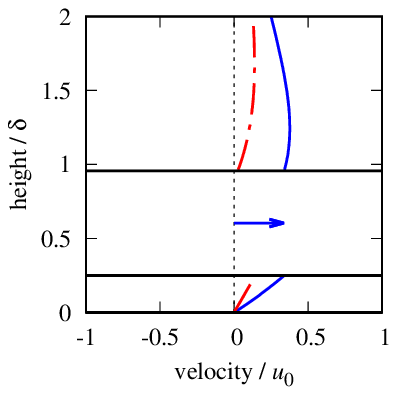}
      \caption{\label{velocity_profile.theta_pio2.d_0.25}
               $d = \delta/4$, $\theta = \pi/2$, $\Im(\sigma)>0$, $\Delta f < 0$}
    \end{subfigure}
  \end{center}
  \caption{\label{velocity_profiles}Velocity profiles for two different
  distances between the plate and the oscillating lower plane (\textit{top row}:
  $d = \delta$, \textit{bottom row}: $d = \delta/4$). The left column shows the
  velocities at a $\theta = \omega t = 0\ \mathrm{rad}$ phase angle, and the
  right column corresponds to a phase angle of
  $\theta = \omega t = \pi/2\ \mathrm{rad}$. The arrows indicate the velocity of
  the plate. The solid blue line plots the velocity of the fluid at different
  heights and the dashed-dotted red line indicates the perturbation of the
  Stokes flow $u_p$ from Eq. (\ref{perturbation_of_flow}) due to the presence of
  the immersed plate.}
\end{figure}

To understand why the  hydrodynamic perturbation of the analyte may produce a positive
frequency shift (or equivalently a ``negative acoustic mass''), let us consider the
tangential hydrodynamic stress  at the surface. Without loss of generality, suppose
the resonator is moving with a velocity $u_0 \cos(\omega t)$, with $u_0$ a real number.
The observed stress $\tilde{\sigma}(t)$ can be decomposed into in-phase and
out-of-phase components, proportional to the real and imaginary parts of the stress
phasor, $\tilde{\sigma}(t) = \Re\left(\sigma e^{-\mathrm{i} \omega t}\right)$.
Now, for phase angles $\theta = \omega t$ equal to integer multiples of $2\pi$, the
resonator velocity reaches its maximum value, $|u_0|$, as its displacement crosses the
midpoint of the oscillation. At this precise moment, the observed stress equals
$\tilde{\sigma}(2 \pi n / \omega) = \Re(\sigma)$, revealing the dissipative part of
the stress. A quarter of a cycle later, ($\theta = 2 \pi n + \pi/2$), the observed
stress equals $\tilde{\sigma}((2 \pi n + \pi / 2)/\omega) = \Im(\sigma)$, which unveils
the fate of the frequency shift. If $\Im(\sigma) > 0$ the extra stress created by the
analyte tends to pull the resonator forward (along $x > 0$), thus decreasing its frequency (negative acoustic mass). The opposite change takes place when
$\Im(\sigma) < 0$. In other words, the ``acoustic mass'' or the frequency shift
simply emerges from the phase lag between the resonator velocity and the extra stress
coming from the analyte. This phase lag is proportional to the time required by
viscous diffusion to  propagate the surface stress from the plate at $z = d$ to the
wall at $z = 0$.

To visualise the impedance in terms of the flow, consider the velocity profiles
drawn in Fig. \ref{velocity_profiles}. The red dashed-dotted lines
correspond to the perturbation $\tilde{u}_p(z, t)$ of the laminar Stokes flow
(\ref{Stokes_boundary_layer}) due to the presence of the immersed plate.
\begin{equation}
  \label{perturbation_of_flow}
  \tilde{u}_p(z, t)
    = \tilde{u}_j(z, t) - \left( u_0 e^{-\alpha z} \right) e^{-\mathrm{i} \omega t},
\end{equation}
where $j$ equals 1 or 2 depending on whether we focus on the fluid below the
plate ($z < d$) or above it ($z > d + a$). Because the extra hydrodynamic stress
caused by the plate is $\tilde{\sigma} =\eta \partial_z \tilde{u}_p$, the real part
of $Z_L$ is proportional to the derivative of $\tilde{u}_p$ with $z$ at a phase angle
of $\theta = \omega t = 0\ \mathrm{rad}$, while the imaginary part corresponds to
the derivative at phase angle $\theta = \pi/2\ \mathrm{rad}$. In the figure, the
sign of $Z_L$ and the surface stress $\tilde{\sigma}$ depends on the slope of the red
dashed-dotted line representing $\tilde{u}_p$ with respect to the vertical dotted line,
which stands for no perturbation ($\tilde{\sigma} = 0$). Notice that at
$\theta = \pi/2$, the slope at $z = 0$ in the top right figure has a sign opposite to
that of the bottom right one, indicating the change in the imaginary part of $Z_L$.
The top panel corresponds to a positive frequency shift $\Delta f$, while the botom
panel yields $\Delta f < 0$. In other words, the top row leads to a ``negative acoustic
mass'', while the bottom row produces a positive result.

\subsection{\label{plate_vs_sphere}Comparing immersed plates with suspended spheres}

\begin{figure}
  \begin{center}
    \begin{subfigure}[b]{.485\linewidth}
      \includegraphics[width = \linewidth]{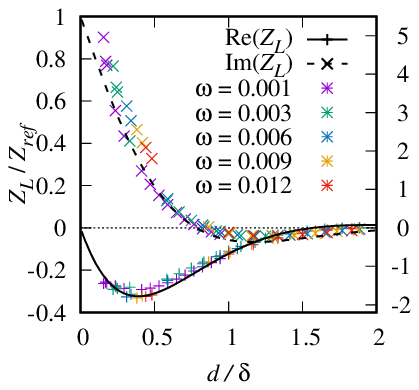}
      \caption{\label{impedance_vs_height.simulations}}
    \end{subfigure}
    \begin{subfigure}[b]{.485\linewidth}
      \includegraphics[width = \linewidth]{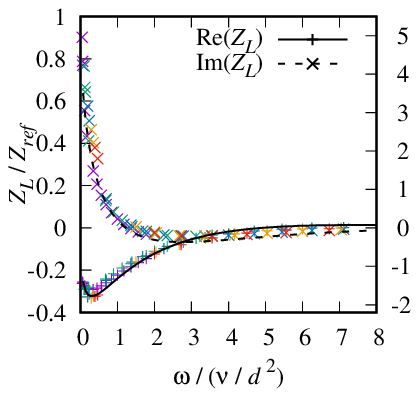}
      \caption{\label{impedance_vs_freq.comparison}}
    \end{subfigure}
    \begin{subfigure}[b]{.485\linewidth}
      \includegraphics[width = \linewidth]{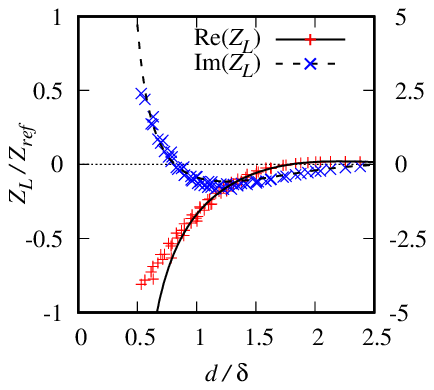}
      \caption{\label{impedance_vs_height.suspended_sphere}}
    \end{subfigure}    
    \begin{subfigure}[b]{.485\linewidth}
      \includegraphics[width = \linewidth]{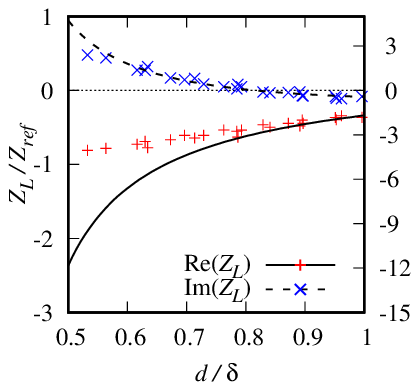}
      \caption{\label{impedance_vs_height.suspended_sphere_close_up}}
    \end{subfigure}
  \end{center}

  \caption{(\subref{impedance_vs_height.simulations}) Load impedance obtained
           from numerical simulations of a suspended neutrally-buoyant sphere
           of radius $R = 0.16\ \delta$ in a
           $(L \times L \times L_z)=(1.33 \times 1.33 \times 5.34) \delta^3$
           box with periodic boundaries in the $x$ and $y$ directions
           versus the height of its centre (points, impedance scale on the right
           axis), compared to the impedance of a plate (curves, left axis) versus
           its distance to the oscillating plane. Results are scaled with the
           Sauerbrey impedance, $Z_{ref}= m \omega$ where the masses per unit
           surface $m = (4\pi/3) R^3\rho'/L^2$ (sphere) and $m = \rho' a$ (plate)
           were chosen equal to each other. The data was obtained from simulations
           at different frequencies.
           (\subref{impedance_vs_freq.comparison}) Load impedance versus
           frequency for the small sphere in
           Fig. \ref{impedance_vs_height.simulations}. Points represent 
           simulation data (right axis), and curves represent the analytical
           result for the solid plate (left axis).
           (\subref{impedance_vs_height.suspended_sphere}) Impedance of a sphere with
           radius $R = 0.526\ \delta$ (points corresponding to the right axis)
           as a function of the distance $d$ between its centre and the wall.
           The curves correspond to the left axis and show the
           impedance caused by a plane at height $d$ with the same lateral 
           motion as a sphere. (\subref{impedance_vs_height.suspended_sphere_close_up}) Close up of Fig. \ref{impedance_vs_height.suspended_sphere} for simulations
           with the sphere close to the wall.}
\end{figure}

Comparing the impedance curve for a solid plate to simulation data for
three-dimensional suspended spheres leads to some interesting and surprising
observations. A few words concering the setup are first in order. These simulations
were performed using the immersed boundary method \cite{Peskin2002,Balboa2012} with
periodic boundary conditions in the $x$ and $y$ directions (resonator plane) and
introducing no-slip rigid planes at the top and bottom of the simulation box. The
oscillating flow was imposed at the bottom of the box and the velocity was set to
zero at the top. The analytical expression for the contribution of the upper boundary
to the impedance results from Eq. (\ref{load_impedance}), setting $v_0 = 0$ and
$d$ equal to the box height.  Both theory and simulations confirm that the change in
the impedance due to a stationary upper wall remains negligible when the box height
exceeds about $3\ \delta$.

Figures \ref{impedance_vs_height.simulations} and \ref{impedance_vs_freq.comparison}
illustrate the proportionality between the impedance due to a small sphere
($R = 0.16\ \delta$ in the figures) and that of a rigid plate at the same height as
the centre of the sphere. The parallel behaviours remain similar up to distances
surprisingly close to the wall.

A large sphere qualitatively changes the behaviour of the impedance near the wall.
While an immersed plate feels the effect of the flow only at height
$d$ (remember that the thickness of the plate plays no role as long as we fix
the value of $\rho' a$), the drag on the sphere comes from the different flow
velocities in the range $z \in [d - R,\ d + R]$, which we can only neglect when
$R \ll \delta$. A simple way to approximate the response of a sphere with this 
one-dimensional model, though, consists in forcing the immersed plate to move in such
a way that its lateral displacement mirrors that of a sphere of radius $R$ at height
$d$ in response to the oscillating flow.

In the steady state, the sphere vibrates with frequency $\omega$,
\begin{equation}
  x(t) = x_0 e^{-\mathrm{i} \omega t}.
\end{equation}
To determine $x_0$, we substitute $x(t)$ into Newton's second law,
\begin{equation}
  \label{Mazur}
  -m\omega^2 x_0 = \mathrm{i} \omega \zeta x_0
                   + 6 \pi \eta r \left[(1 + \alpha r) \bar{v}_s + \frac{1}{3} \alpha^2 r^2 \bar{v}_v \right].
\end{equation}
The force phasor amplitude on the right was calculated by Mazur and Bedeaux in Ref.
\cite{Mazur1974}. The friction $\zeta$ stands for
\begin{equation}
  \zeta = 6 \pi \eta r \left( 1 + \alpha r + \frac{1}{9} \alpha^2 r^2 \right),
\end{equation}
and $\bar{v}_s$ and $\bar{v}_v$ for averages of the unperturbed flow over the
surface and volume of the sphere respectively. Their analytical expressions are
derived in appendix A. Solving Eq. (\ref{Mazur}) for the phasor describing the
motion of the sphere, we get
\begin{equation}
  \label{position_phasor}
  x_0 = \frac{6 \pi \eta r \left[ (1 + \alpha r) \bar{v}_s + \frac{1}{3} \alpha^2 r^2 \bar{v}_v \right]}
             {-m\omega^2 -\mathrm{i} \omega \zeta}.
\end{equation}
The velocity phase amplitude equals $v_0 = -\mathrm{i} \omega x_0$, so the corresponding
load impedance follows from Eq. (\ref{load_impedance}) writing $v_0$ in terms of the
$x_0$ given above. Plotting the impedance for a sphere with a diameter comparable to
the penentration depth, $R/\delta = 0.526$, produces the curves in Fig.
\ref{impedance_vs_height.suspended_sphere}. Once again, apart from the vertical scaling
factor, the curves agree as the sphere moves away from the wall, even though we are
comparing its impedance to that of a plane.

Eq. (\ref{Mazur}) works well far from the wall but breaks down close to it. As Mazur
and Bedeaux themselves pointed out \cite{Mazur1974}, the theory does not take into
account the hydrodynamic reflections that significantly modify the Stokes flow felt by
the sphere when it approaches the resonator surface. Fig. 
\ref{impedance_vs_height.suspended_sphere_close_up}  confirms that the approximate
theory and simulations significantly disagree near the oscillating wall.

\begin{figure}
  \begin{center}
    \includegraphics[width =  \linewidth]{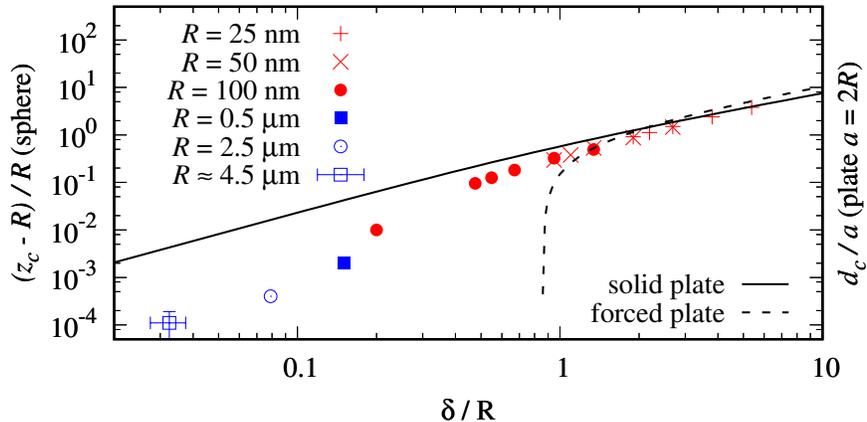}
  \end{center}
  \caption{\label{dcut_vs_delta}Scaled zero-frequency-shift separation for particles
  (\textit{points}) and plates (\textit{lines}) versus scaled penetration depth. The
  points correspond to simulations (\textit{red}) and experiments (\textit{blue}) for
  particles of different sizes ($R = 0.5\ \mu\textrm{m}$ and $R = 2.5\ \mu\textrm{m}$
  from Ref. \cite{Olsson2012}, and $R = 2\ \mu\textrm{m}$ from Refs.
  \cite{Pomorska2010,Pomorska2011}). The solid line represents the plate model from Eq. (\ref{solid_plate_load_impedance}), while the dashed line plots the forced plate
  model from section \ref{plate_vs_sphere}.}
\end{figure}

The crossing over to positive frequency shifts ($\Im(Z_L) < 0$) has received attention
in experimental research, where it has been viewed as a proxy for interactions between
large particles and the substrate
\cite{Dybwad1985,Pomorska2010,Olsson2012,Johannsmann2015}. Fig. \ref{dcut_vs_delta}
presents the scaled crossover separation between particles and QCM surface as a
function of the scaled penetration depth. Rescaling the particle radius $R$ and the
penetration depth $\delta$ by the same factor results in an equivalent flow. Therefore,
when we represent the zeros, $z_c$, of $\Im(Z_L) = 0$ divided by $R$ versus $\delta/R$
for simulations of different spheres and penetration depths, they all collapse onto the
same curve. The penetration depth contains the frequency dependence of $z_c$ with
respect to $\omega$ because $\delta \propto \omega^{-1/2}$. The solid line follows the
plate model prediction of Eq. (\ref{crossover_distance}) for $a = 2R$, while the
dashed line corresponds to the prediction of the forced plate model in this section (Eqs.
(\ref{load_impedance}) and (\ref{position_phasor}), setting $\Im(Z_L) = 0$). The latter
model gives an indication of how the acoustic response changes with the sphere dynamics,
which arise from the forces induced by the surrounding flow. For large values of
$\delta/R$ (low frequencies or small particles) we observe similar crossover values
for the plates and particles. Clearly, the theories for plates depart from the behaviour
of spheres when the penetration depth becomes comparable to the sphere radius, with
significant deviations when $\delta/R < 1.25$. For a QCM frequency of $35\ \mathrm{MHz}$
this corresponds to spheres with $R > 50\ \mathrm{nm}$. Close to this penetration depth
$\delta/R \approx 1.25$, the forced plate model yields slightly better predictions,
suggesting that the crossover height decreases more quickly than in the solid plate model
due to the sphere dynamics. However, as $\delta/R$ is further decreased, the forced plate
model largely overestimates the decay of the cross-over distance. When the spheres lie
close to the resonator ($\delta/R < 1.25$), multiple hydrodynamic reflections between the
resonator and the particle determine the flow and hydrodynamic impedance.

Fig. \ref{dcut_vs_delta} also includes some experimental observations from Pomorska
\textit{et al.} \cite{Pomorska2010} and Olsson \textit{et al.} \cite{Olsson2012} (in
blue). Let us first turn our attention to the latter reference. There, the authors
observed silica particles over a bare silica surface in a (1:1) electrolite
($\mathrm{K}^{+}$, $\mathrm{Cl^{+}}$) at different ionic strengths (from 0 to 150 mM).
Metallurgical microscopy determined that the particles performed Brownian motion above
the surface. Adding enough electrolite ($c=150\mathrm{mM}$) reduced the Brownian motion,
indicating the screening of repulsive electrostatic forces and adsorption by
dispersion (van der Waals) forces. Although not explictly mentioned in \cite{Olsson2012},
at smaller ionic strengths one expects to find the silica particles
suspended over the resonator and exposed to the wall-interaction potential.
According to the DLVO theory, at low ionic strengths, below the critical coagulation
concentration, $c_{ccc}$, this potential has a secondary minimum at a distance
of about $d \approx 6/\kappa$ ($\kappa$ stands for the Debye-Hückel screening length)
\cite{Israelachvili2010}. For $c>c_{ccc}$, the particles start to adhere to the surface
due to dispersion forces. For a KCl electrolite in water, the Debye length 
($\kappa^{-1} \propto c^{-1/2}$) is about $10\ \mathrm{nm}$ for
$c \approx 1\ \mathrm{mM}$. Taking the values of the crossover frequency $\omega_c$
reported by Olsson \textit{et al.} \cite{Olsson2012} (which grow with the ionic
strength), we can extrapolate the the tendency observed in our simulations to estimate
the typical distance $d$ between silica partices and the surface. Notably, the result
of this crude estimation agrees with distances $d$ decreasing with $c$ as $d\sim
6/\kappa$, which points to an acoustic response governed by hydrodynamics in these
experiments. A quantititative prediction would require an elaborate theory (which
should weight the impedance-height dependence) including a more complete set of
experimental details (surface charge values, for example). We have recently carried
out a detailed analysis in the case of suspended liposomes tethered to DNA strands
\cite{Delgado2020}. The close agreement between experiments and simulations confirmed
the dominant role of the hydrodynamic impedance when dealing with suspended particles,
and enabled quantitative predictions.

A second set of experiments by Olsson \textit{et al.} \cite{Olsson2012} considered
streptavidin-decorated silica particles adsorbed to a biotinylated silica surface.
In that case the strong streptavidin-biotin links gradually adsorbed the particles
and the results for the crossover frequency vary only mildly with the ionic strength.
The typical particle-surface distance $d$ corresponds to molecular contact
(1 nm or less). Figure \ref{dcut_vs_delta} shows that this estimation
is also compatible with our hydrodynamic predition.

The experiments by Pomorska {\em et al.} \cite{Pomorska2010} provide further evidence
of the importance of hydrodynamics, even when particles are adsorbed. In those
experiments the particles and surface were decorated with polyelectrolites of
opposite charge to ensure a strong attractive potential and adhesion.
The size of the particles in these experiments was about $4.5\ \mu\mathrm{m}$ and
the crossover frequency was close to $15\ \mathrm{MHz}$ for the two cases considered.
We set the distance to the resonator equal to a molecular contact
($d_c \in [0.5-1]\ \mathrm{nm}$) to plot the experimental data in Fig.
\ref{dcut_vs_delta} (blue squares). The point nicely extrapolates the trend we
predicted for much smaller particles.

In summary, our analyses provide evidence that the leading contribution to the
load impedance created by analytes immersed in liquids comes from hydrodynamics.
While we have recently proved this claim in the case of suspended particles (liposomes
tethered to DNA \cite{Delgado2020}), in the case of adsorbed particles, our findings
call for a revision of the relevance and estimation of contact forces from QCM analyses.

\section{Elastic layer}

\begin{figure}
  \begin{center}
    \begin{subfigure}[b]{.485\linewidth}
      \includegraphics[width = \linewidth]{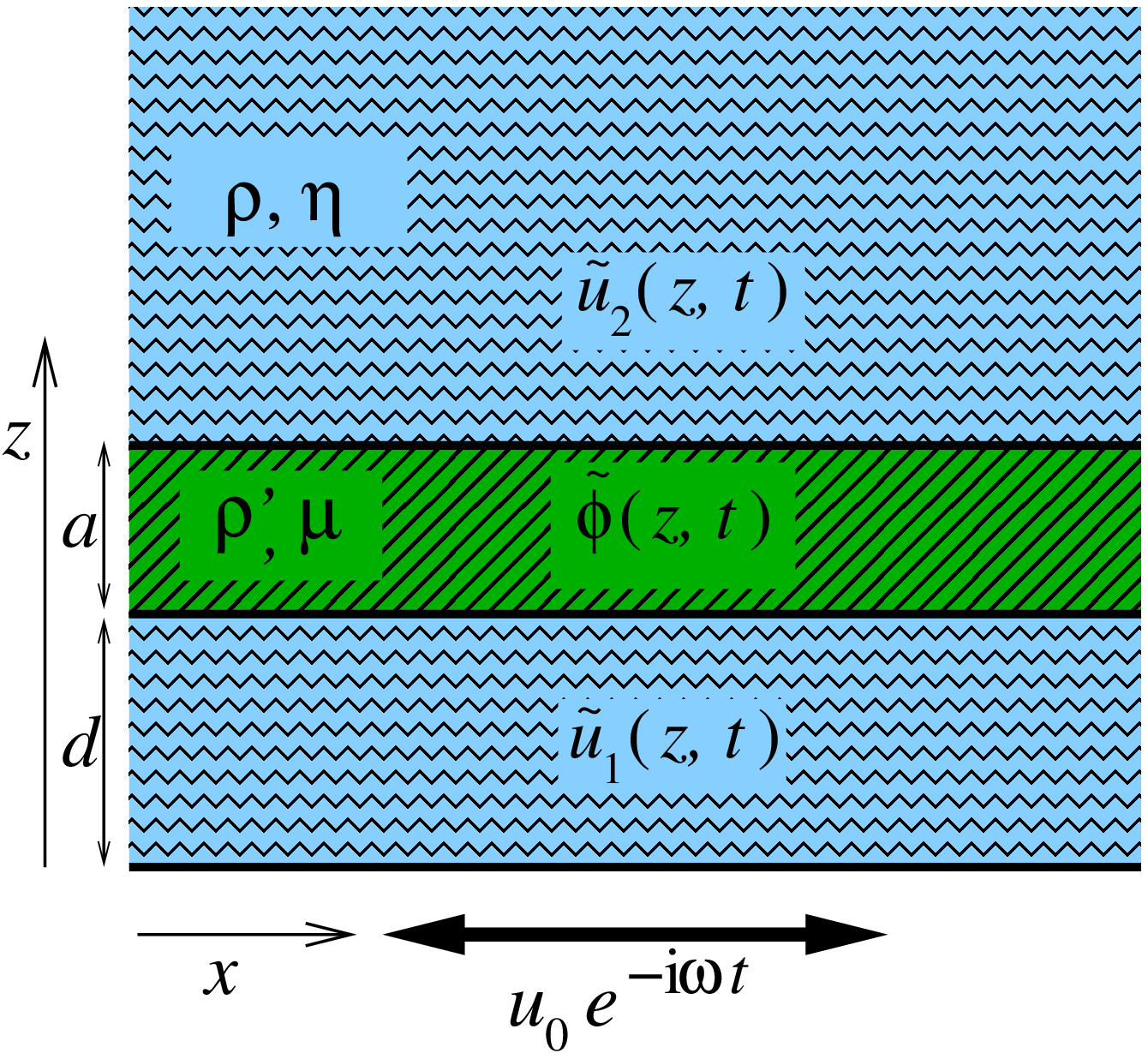}
      \caption{\label{elastic_layer_diagram}}
    \end{subfigure}
    \begin{subfigure}[b]{.485\linewidth}
      \includegraphics[width = \linewidth]{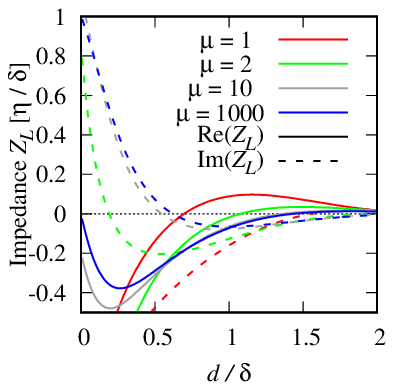}
      \caption{\label{impedance_vs_height.elastic_layer}}
    \end{subfigure}
    \begin{subfigure}[b]{\linewidth}
      \includegraphics[width = \linewidth]{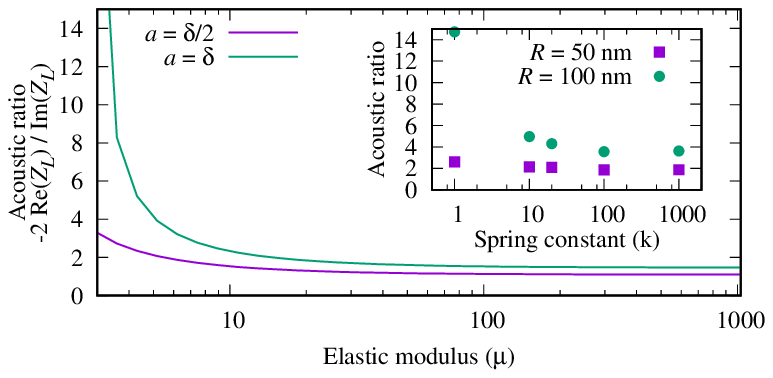}
      \caption{\label{AR_vs_klipo}}
    \end{subfigure}
  \end{center}
  \caption{(\subref{elastic_layer_diagram}) Simple model of an elastic layer,
           where an elastic material of density $\rho'$ and shear modulus $\mu$
           replaces the plate. The function $\tilde{\phi}(z, t)$ indicates the
           displacement in the $x$ direction at height $z$ and time $t$ within
           the layer.
           (\subref{impedance_vs_height.elastic_layer}) Impedance due to an
           elastic layer of density $\rho' = \rho$ and thickness
           $a = \delta/\sqrt{2}$ as a function of the distance $d$ to the lower
           plane. As the material becomes more rigid, the curves approach the
           solution for the solid plate (compare the curve for $\mu = 1000$ to
           the green line for $\rho' = \rho$ in Fig. \ref{impedance_vs_height}).
           (\subref{AR_vs_klipo}) Acoustic ratio of a neutrally-buoyant elastic
           layer of thickness $a$ versus shear modulus $\mu$ at
           $d = 0.18\ \delta$, compared to simulations of spherical liposomes of
           radius $R$ at height $d + R$ as a function of the elastic strength of the
           bonds used to connect neighbouring elements in the numerical model
           (\textit{inset}).}
\end{figure}

Let us replace the solid plate with an elastic layer of thickness $a$,
density $\rho'$ and shear modulus $\mu$ (Fig. \ref{elastic_layer_diagram}).
Within the layer, we denote the displacement of a point at height $z$ and
time $t$ along the $x$ direction with $\tilde{\phi}(z, t)$, which must satisfy the
equation of motion \cite{Kanazawa1985}
\begin{equation*}
  \frac{\partial^2 \tilde{\phi}(z, t)}{\partial t^2}
    = \frac{\mu}{\rho'} \frac{\partial^2 \tilde{\phi}(z, t)}{\partial z^2},
\end{equation*}
a wave equation with speed $c = \sqrt{\mu/\rho'}$. The steady state solution at
frequency $\omega$ equals
\begin{equation*}
  \tilde{\phi}(z, t)
    = \Re\left(\left(C e^{-\mathrm{i} k z} + D e^{\mathrm{i} k z}\right)
               e^{-\mathrm{i} \omega t}\right),
\end{equation*}
with $k = \omega/c$. Imposing the boundary conditions on $\tilde{u}_1$, $\tilde{u}_2$
and $\tilde{\phi}$, which amount to continuity in the speeds and stresses plus the no
slip condition at $z = 0$ and vanishing velocity as $z$ tends towards infinity,
\begin{align*}
  \tilde{u}_1(0, t) & = u_0 e^{-\mathrm{i} \omega t}, \\
  \tilde{u}_1(d, t)
    & = \left. \frac{\partial \tilde{\phi}}{\partial t} \right|_{z = d}, \\
  \tilde{u}_2(d + a, t)
    & = \left. \frac{\partial \tilde{\phi}}{\partial t} \right|_{z = d + a}, \\
  \lim_{z \to \infty} \tilde{u}_2(z, t) & = 0, \\
  \eta \left. \frac{\partial \tilde{u}_1}{\partial z} \right|_{z = d}
    & = \mu \left. \frac{\partial \tilde{\phi}}{\partial z} \right|_{z = d}, \\
  \eta \left. \frac{\partial \tilde{u}_2}{\partial z} \right|_{z = d + a}
    & = \mu \left. \frac{\partial \tilde{\phi}}{\partial z} \right|_{z = d + a},
\end{align*}
we obtain with the help of some computer algebra the following value for the
load impedance,
\begin{equation*}
  Z_L = \frac{2 \alpha \eta (1 - \Lambda^2)
              \left(e^{2 \mathrm{i} a \omega / c} - 1\right)}
             {e^{2 \alpha d}
              \left((1 + \Lambda)^2
                    - e^{2 \mathrm{i} a \omega / c} (1 - \Lambda)^2\right)
              + (1 - \Lambda^2)\left(e^{2 \mathrm{i} a \omega / c} - 1\right)}.
\end{equation*}
$\Lambda$ represents the dimensionless parameter
\begin{equation}
  \Lambda = \frac{\alpha \eta}{\rho' c},
\end{equation}
proportional to the ratio of the velocity of viscous diffusion over $\delta$ to
the speed of elastic waves, $c$. The other relevant groups are phase lags, $a\omega/c$
and $\alpha d$.

When the elastic medium becomes rigid ($c \to \infty$, $\Lambda \to 0$) we recover
the solution for the rigid plate. Figure \ref{impedance_vs_height.elastic_layer}
plots the impedance due to the layer as a function of the distance $d$ that separates
it from the vibrating plane.

Using the computational methods mentioned in the previous section, we simulated
elastic neutrally-buoyant liposomes using an elastic network made up of elements
connected by harmonic bonds of spring constant $k$. We observed that the acoustic ratio
decreased as we increased $k$. Increasing the rigidity of our layer leads to similar
predictions (see Fig. \ref{AR_vs_klipo}).

\section{Fluid layer}

\begin{figure}
  \begin{center}
    \begin{subfigure}[b]{.485\linewidth}
      \includegraphics[width = \linewidth]{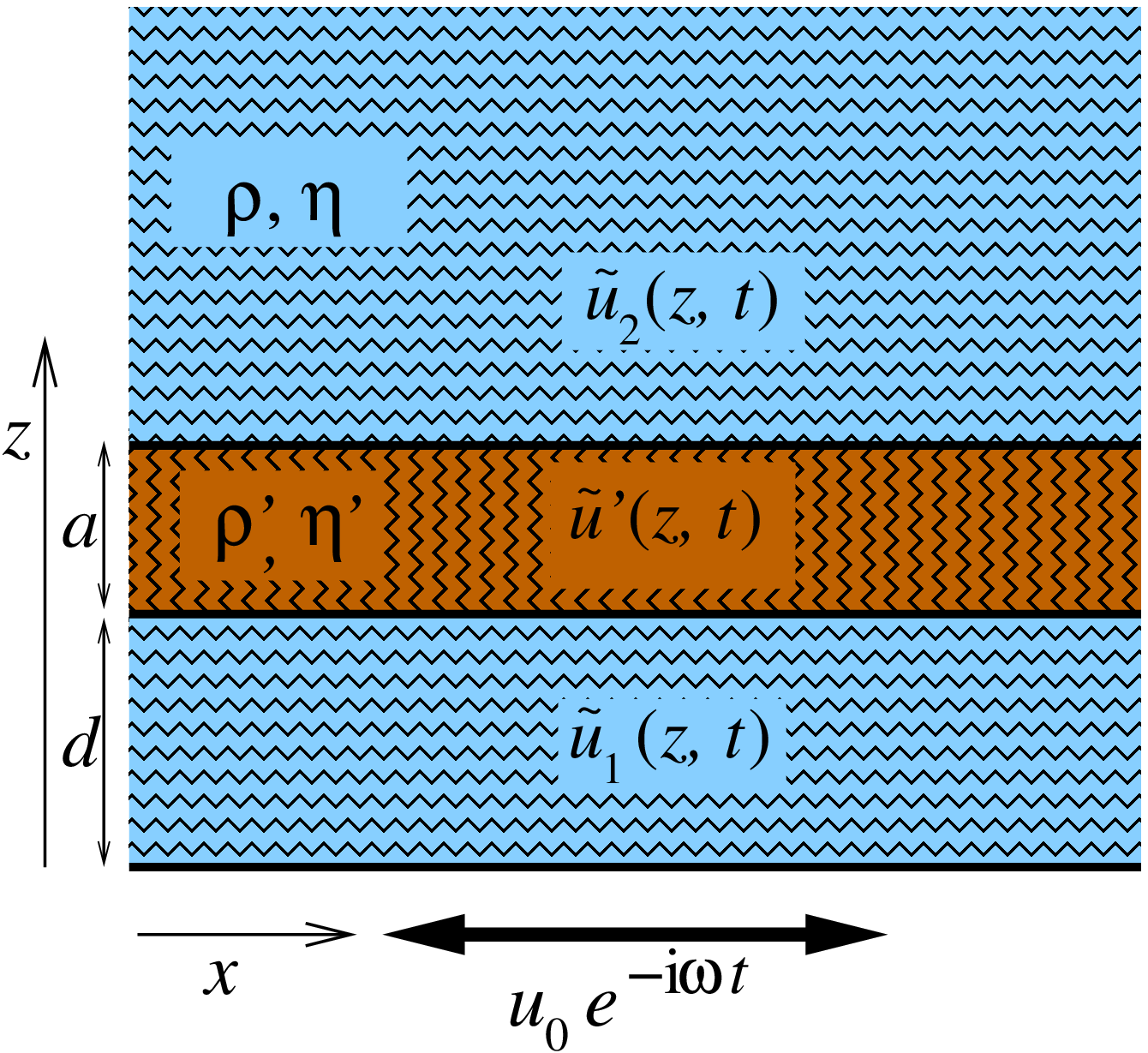}
      \caption{\label{fluid_layer_diagram}}
    \end{subfigure}
    \begin{subfigure}[b]{.485\linewidth}
      \includegraphics[width = \linewidth]{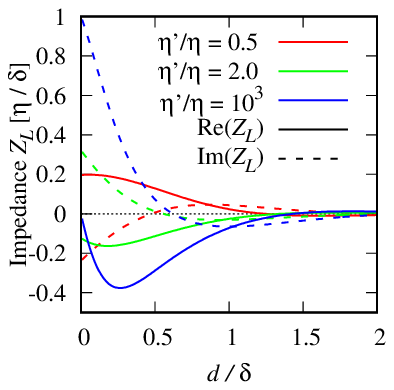}
      \caption{\label{impedance_vs_height.fluid_layer}}
    \end{subfigure}
  \end{center}
  \caption{(\subref{fluid_layer_diagram}) Simple model of a fluid layer, with
           a liquid of density $\rho'$ and shear viscosity $\eta'$ instead of a
           solid layer. The function $u'(z, t)$ names the velocity field inside
           the layer.
           (\subref{impedance_vs_height.fluid_layer}) Impedance due to a
           fluid layer of density $\rho' = \rho$ and thickness 
           $a = \delta/\sqrt{2}$ as a function of the distance $d$ to the lower
           plane. As the viscosity increases, the curves look more and more like
           those of the solid plate (the curves for $\eta'/\eta = 10^3$ resemble
           the green lines for $\rho' = \rho$ in Fig.
           \ref{impedance_vs_height}).}
\end{figure}

Lastly, we will work out the impedance for a plane fluid layer of density
$\rho'$, shear viscosity $\eta'$ and velocity field $u'(z, t)$. Fig.
\ref{fluid_layer_diagram} displays a sketch of the system. Once again, we
express the fluid velocities with Eq. (\ref{Stokes_boundary_layer}) and impose
the appropriate boundary conditions (continuity of velocities and stress, no
slip at the lower boundary and vanishing velocity as $z$ tends towards
infinity). The resulting load impedance equals
\begin{equation}
  Z_L = 
    \frac{2 \alpha \eta e^{-2 \alpha d}\tanh(\alpha' a)(1 - \Upsilon^2)}
         {\tanh(\alpha' a) \left(1 + e^{-2 \alpha d}
                                 + \Upsilon^2 \left(1 - e^{-2 \alpha d}\right)
                           \right) + 2 \Upsilon},
\end{equation}
with
\begin{equation}
  \alpha' = (1 - \mathrm{i}) \sqrt{\frac{\omega \rho'}{2 \eta'}},
\end{equation}
and
\begin{equation}
    \Upsilon = \frac{\alpha' \eta'}{\alpha \eta}.
\end{equation}
Figure \ref{impedance_vs_height.fluid_layer} shows the change in the
impedance of the fluid layer as a function of the distance $d$ to the lower
plane for different values of the shear viscosity. As the viscosity increases,
the curves approach the solid plate limit. Interestingly, a layer viscosity
lower than that of the surrounding fluid leads to a flip in the behaviour of the
real and imaginary parts of $Z_L$, as observed in QCM experiments with nanobubbles
\cite{Du2004,Zhang2008,Ondarcuhu2013}
(in the figure, compare the red line for $\eta'/\eta = 0.5$ to the green line for $\eta'/\eta = 2$).

\section{Discussion}

In addition to providing analytical expressions for the load impedance of
different types of immersed layers, we have demonstrated the importance of
considering the role of hydrodynamics in explaining the effects of these
layers on the QCM. Although we have not considered any contact forces between
the load and the QCM, the models explained above predict the behaviour of
suspended loads and recover the expected Sauerbrey relation in the limit
of adsorbed layers. Furthermore, the ``vanishing mass'' phenomenon observed in
suspensions arises as a natural consequence in our derivations.

\begin{figure}
  \begin{center}
    \includegraphics[width =  \linewidth]{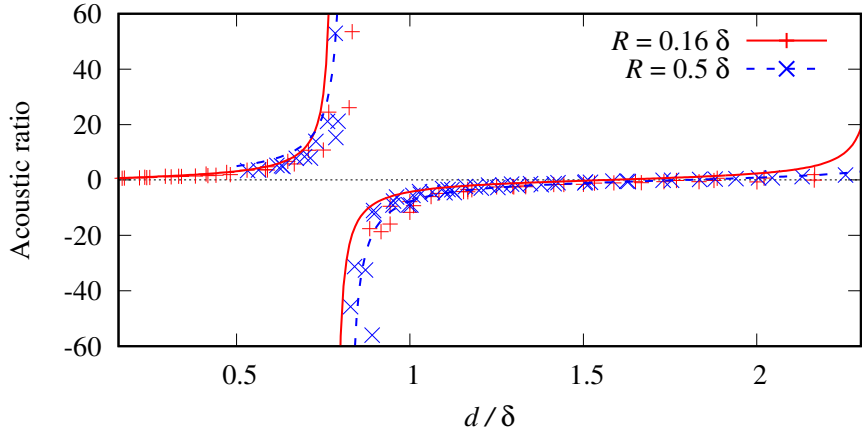}
  \end{center}
  \caption{\label{AR_vs_height.suspended_sphere}Acoustic ratio vs height overt the
  QCM for neutrally-buoyant spheres (points) and solid plate (lines). The distance
  $d$ indicates the separation between the plate and QCM, and the distance from the
  center of the sphere to the QCM.}
\end{figure}

The evidence provided here strongly suggests that other types of suspensions
(such as the simulated suspended spheres considered above) share the same
generic features. Even though the plates in section \ref{plate_vs_sphere} had an
impedance about five times greater than the spheres, the dependence on height
displayed surprisingly parallel behaviours. Prefactors cancel out when calculating
the acoustic ratio, so the plate acoustic ratios provide a decent estimate of the
value measured for sphere (see Fig. \ref{AR_vs_height.suspended_sphere}).

The preceding pages show that large acoustic ratios do not necessarily imply large
values of the dissipation. As we have seen, vanishing frequency shifts naturally lead
to diverging acoustic ratios.

Finally, we considered the crossover to positive frequency shifts and compared the analytical prediction of the one-dimensional plate system to simulations of
sub-micron spheres and experiments carried out with micron-sized colloids.  The
plate model correctly predicts the zero-frequency crossover for small enough
particles. We observe a transition to a large particle regime when the dimensionless
parameter $R/\delta$ becomes large enough ($R/\delta > 0.8$). The one-dimensional
theories clearly fail in this regime. By contrast, our three-dimensional simulations
correctly extrapolate to larger (micron-sized) colloids, even with the latter
adsorbed to the wall. As we did not consider adhesive forces, such an agreement 
highlights the role of hydrodynamics in determining the response of large adsorbed
particles, and calls for a hydrodynamic extension of the existing contact-force and
elastic-stiffness QCM theories. The ``coupled-resonance model'', which predicts
positive shifts within the ``elastic loading'' regime in QCM
\cite{Dybwad1985,Johannsmann2015} was originally derived for spheres in the dry state
but has subsequently been applied extensively in liquids
\cite{Pomorska2010,Olsson2012,Johannsmann2015}. Our results call for a revision of the
role of contact forces and elastic stiffness in liquids, an investigation which
requires a generalization of existing theories including the difficult topic
of elastohydrodynamic lubrication \cite{Gohar2001}. Advancing our theoretical
undestanding in this direction would greatly improve  the  predictive power of QCM
analyses of molecular and mesoscopic contact forces.

\section{Acknowledgements}

Our research here was supported by the European Commission FETOPEN Horizon 2020
Catch-U-DNA project.

\section*{Appendix A. Average velocity integrals}

The velocities $\bar{v}_s$ and $\bar{v}_v$ in Eq. (\ref{Mazur}) come from
averaging the flow profile $u(z, t)$ over the surface and volume of the sphere,
respectively,
\begin{align*}
  \bar{v}_s & = \frac{1}{4 \pi r^2} \int_S u_0 e^{-\alpha z}\ d\mathbf{S}, \\
  \bar{v}_v & = \frac{3}{4 \pi r^3} \int_V u_0 e^{-\alpha z}\ d\mathbf{V}.
\end{align*}
To carry out the first of these integrals, we choose spherical coordinates
around the centre of the immersed sphere at height $d$. Therefore,
\begin{equation*}
  \int_S u(z, t)\ d\mathbf{S}
    = \int_0^{2\pi} \left( \int_0^\pi u_0 e^{-\alpha (d + r\ \cos(\theta))} r^2 \sin(\theta)\ d\theta \right)\ d\phi.
\end{equation*}
After integrating over $\phi$,
\begin{equation*}
  \int_S u(z, t)\ d\mathbf{S}
    = 2 \pi u_0 r e^{-\alpha d} \int_0^\pi e^{-\alpha r\ \cos(\theta)} r\ \sin(\theta)\ d\theta
    = 2 \pi u_0 r e^{-\alpha d} \left[ \frac{e^{-\alpha r\ \cos(\theta)}}{\alpha} \right]_0^\pi.
\end{equation*}
Hence,
\begin{equation*}
  \bar{v}_s = \frac{u_0 e^{-\alpha d}}{\alpha r}\sinh(\alpha r).
\end{equation*}
The volume integral is simply equal to the integral over the radius of the
surface integral from 0 to the radius of the sphere $r$,
\begin{equation*}
  \int_V u(z, t)\ d\mathbf{V}
    = \int_0^r \frac{4 \pi r' u_0}{\alpha} e^{-\alpha d} \sinh(\alpha r') dr'
    = \frac{4 \pi u_0}{\alpha} e^{-\alpha d}
      \left[ \frac{r'\ \cosh(\alpha r')}{\alpha} - \frac{\sinh(\alpha r')}{\alpha^2} \right]_0^r,
\end{equation*}
from which we get
\begin{equation*}
  \bar{v}_v = \frac{3 u_0}{\alpha r^3} e^{-\alpha d}
              \left( \frac{r\ \cosh(\alpha r)}{\alpha} - \frac{\sinh(\alpha r)}{\alpha^2} \right).
\end{equation*}


\end{document}